\documentclass[agupp]{aguplus}

\sectionnumbers
\printfigures

\doublecaption{35pc}

\authoraddr{Corresponding author: Dr. Helmut Z. Baumert,\\ IAMARIS,  B\"urgermeister-Jantzen-Str. 3,
19288 Ludwigslust, Germany (baumert@iamaris.org)}

\slugcomment{Submitted for publication to {\it Physica Scripta} (IOP): 29 March, 2015}

\usepackage{graphics}
\usepackage{graphicx}
\graphicspath{%
    {converted_graphics/}
    {/}
}
\begin{document}
\bibliographystyle{ametsoc}

\def\lesssim{\mathrel{\hbox{\rlap{\hbox{\lower0.45em\hbox{$\sim$}}}\hbox{$<$}}}}
\def\gtrsim{\mathrel{\hbox{\rlap{\hbox{\lower0.45em\hbox{$\sim$}}}\hbox{$>$}}}}
\def\gae{\mathrel{\hbox{\rlap{\hbox{\lower0.45em\hbox{$\sim$}}}\hbox{$>$}}}}
\def\lae{\mathrel{\hbox{\rlap{\hbox{\lower0.45em\hbox{$\sim$}}}\hbox{$<$}}}}

\title{
Turbulent mixing and a  generalized\\ phase transition in shear-thickening fluids  }

\author{Helmut Z.\ Baumert}
\affil{IAMARIS, Ludwigslust, Germany\\ (baumert@iamaris.org)}

\author{Bernhard Wessling}
\affil{LaoWei Chemical Technology Consulting Inc., Shenzhen, China  (wessling@LaoWei-Consulting.com)}

\begin{abstract}
\noindent
This paper presents a  new theory of  turbulent mixing in stirred reactors. 
The degree of homogeneity of a mixed fluid may be characterized by (among other features) 
Kolmogorov's microscale, $\lambda$. The smaller its value, the better the homogeneity. 
According to Kolmogorov, $\lambda$ scales inversely with the fourth root of the energy flux applied in 
the stirring process, $\varepsilon$. The higher $\varepsilon$, the smaller $\lambda$, and the better
the homogeneity in the reactor. This is true for Newtonian fluids. 
In non-Newtonian fluids the situation is different. For instance, in shear-thickening fluids 
it is plausible that  high shear rates thicken the fluid and might strangle the mixing. 
The internal interactions between different fluid-mechanical and colloidal variables are subtle, namely
due to the (until recently) very limited understanding of turbulence. 
\newline $\quad$Starting from a qualitatively new turbulence theory for inviscid fluids \cite[]{baumert2013}, giving e.g.\ 
Karman's constant as $1/\sqrt{2\,\pi}= 0.40$  \cite[Princeton's  superpipe gives  0.40$\pm 0.02$,][]{baileyetal2014}, 
we generalize this approach to the case of viscous fluids and derive equations which in the steady state exhibit 
two solutions. One solution branch describes a state of good mixing, the other strangled turbulence. 
The physical system cannot be in two different steady states at the same time. But it is physically admissible 
to switch between the two steady states by non-stationary transitions, maybe in a chaotic fashion.

\vspace{0.25cm}
\noindent
{\bf Keywords:} Turbulence, kinematic viscosity, eddy viscosity, colloidal systems, dispersions,
shear-thickening, mixing, stirring, reactor, non-Newtonian fluid, Kolmogorov scale, bifurcation

\end{abstract}


\section{Introduction\label{intro}}  

\subsection*{Plausibility considerations\label{plausi}}

A mixer is a reactor wherein fluids are mixed with other fluids, with dry or liquid chemicals, heat or other so-called `scalars'.
To that aim propellers, screws, raising bubbles or other technological stirring means and devices are applied. The near-field 
turbulence initiated by them  is characterized by two parameters: a characteristic length scale and a characteristic time scale 
(or, equivalently, a corresponding wavenumber and a corresponding frequency). These two parameters may be tuned independently
by modifying the geometry and the rotation rate of the device (the bubble-release frequency in bubble or convective mixing). 

If the operator  of a mixer wishes a higher degree of homogeneity then he is inclined to increase the energy spend for mixing
by increasing the rotation rate, the geometric parameters of the propeller, or both. For Newtonian fluids this is a reasonable strategy. 
However, in shear-thickening fluids it is plausible that with an increase of the forcing one might contribute to thickening 
which in the extreme strangles mixing until breakdown of turbulence. Here much depends on details, i.e.\ \textit{how} one increases the
stirring power, by geometric means, by the rotation rate, or by both. 

As mixing and dispersion are fundamental technological
steps in nearly all industries and connected with huge financial fluxes if considered for the globe, this circle of questions deserves attention. 

\subsection*{Historical remarks\label{hist}}

Whereas the arts of hydraulics including the phenomenology of eddying flows and vortices are well known at least for 
about $2.4\times 10^3$ years, back to e.g.\ the `hydraulic empires' of  Nofretete and Echnaton, the mathematical science 
of dynamic fluid motions emerged 'just recently': in idealized, purely geometric form by Leonhard Euler (1707 - 1783), 
supplemented somewhat later with frictional aspects by Claude-Louis Navier and George Stokes. The theory of vortices and 
circulation was more explicitely elaborated about 30 years later by Herrmann von Helmholtz and William Thomson (Lord Kelvin), 
and later came Osborne Reynolds, Ludwig Prandtl, Theodore von Karman, Alexander A. Friedmann and Lev V. Keller with their
specific innovations.

A  systematic statistical treatment of turbulence was initiated by Geoffrey I. Taylor and, in form of spectral considerations, by Andrej N. Kolmogorov.
These attempts were more or less singular pieces, kept  loosely together by semi-empirical scaling rules and similarity arguments. 
Great contributions in this respect stem here from Andrej S. Monin, Alexander Obukhov, Rostislav V. Osmidov, Chuck van Atta and Stephen Thorpe. 

With the advent of sufficiently powerful computers the focus moved towards so-called two-equation turbulence models. Major authors are 
David C. Wilcox, George L. Mellor, Tetsuji Yamada, and Wolfgang Rodi. The overwhelming number of turbulence theories in the above sense 
were elaborated for Newtonian fluids (and gases) wherein kinematic viscosity is more or less constant in larger elements of space 
and time and not dependent neither on the state of the fluid motion nor on external forces, nor on time. 

\subsection*{Smart fluids / colloidal systems\label{smart}}

Today so-called smart fluids become increasingly relevant  in various industries, from medicine over small-scale bio-technology to construction 
engineering, waste-water treatment and defence. The central property of a smart fluid is its kinematic viscosity, 
which may explicitely depend on time or not, which may be tuned by shear or light, by electric or magnetic fields like in magneto-rheological fluids.

As their kinematic viscosity is not a constant, smart fluids have always non-Newtonian character. 
They all are at least 2-phase systems, whereas the -- at least one -- dispersed phase is mostly nanoscopic (i.e., less than 100 nm diameter) in size. 
Hence these are colloidal systems, either colloidal dispersions (the dispersed phase being a solid) or emulsions (the dispersions being a liquid). In contrast 
to wide-spread assumptions, such colloidal systems exhibit complex structures. I.e., the dispersed phase is not statistically evenly distributed, but highly structured 
\cite[]{wessling1993}. This is due to the thermodynamically non-equilibrium character of such colloidal dispersions / emulsion systems \cite[]{wessling1991,wessling1995}.

As long as the flow of a non-Newtonian fluid or gas is laminar, the consequences of those external controls 
on the kinematic viscosity remain more or less predictable and no dispersion would be possible 
\cite[]{wessling1995}. This property is going to be changed when the flow becomes critical and eventually turbulent. 

Even the Reynolds number in non-Newtonian fluids looses much of its classical meaning. Traditionally it is defined as the dimensionless ratio of inertial and viscous forces,
\begin{equation}\label{Reynolds}
	Re = \frac{U\times L}{\nu}
\end{equation}
where $U$ [m\,s$^{-1}$] is a velocity scale for the mean flow, $L$ [m] a characteristic length scale (e.g. a distance
from a solid wall or body) and $\nu$ is the kinematic viscosity [m$^2\,$s$^{-1}$]. This means that $Re$ depends linearly on the state of flow 
via the velocity, $U$. However, if also $\nu$ depends on the flow state then things become more difficult. 

Relatively simple cases of smart fluids are the so-called shear-thickening (dilatant) and the shear-thinning (pseudoplastic) fluids. In the former case
 viscosity {\it increases} with increasing shear rate so that the fluid becomes `thicker'. The most prominent example is the fluid in body armor wests 
of policemen. Certain water-sand mixtures may behave in similar form. Prominent examples for pseudoplastic fluids are lava, blood and whipped cream.
Also the notorious quicksand is an example for a shear-thinning material.

In all these cases the turbulent state of those fluids -- if at all -- is understood only in an empirical sense. 
 \begin{figure}[htbp] \label{kinemat}
  \includegraphics[width=7cm,height=7cm,keepaspectratio]{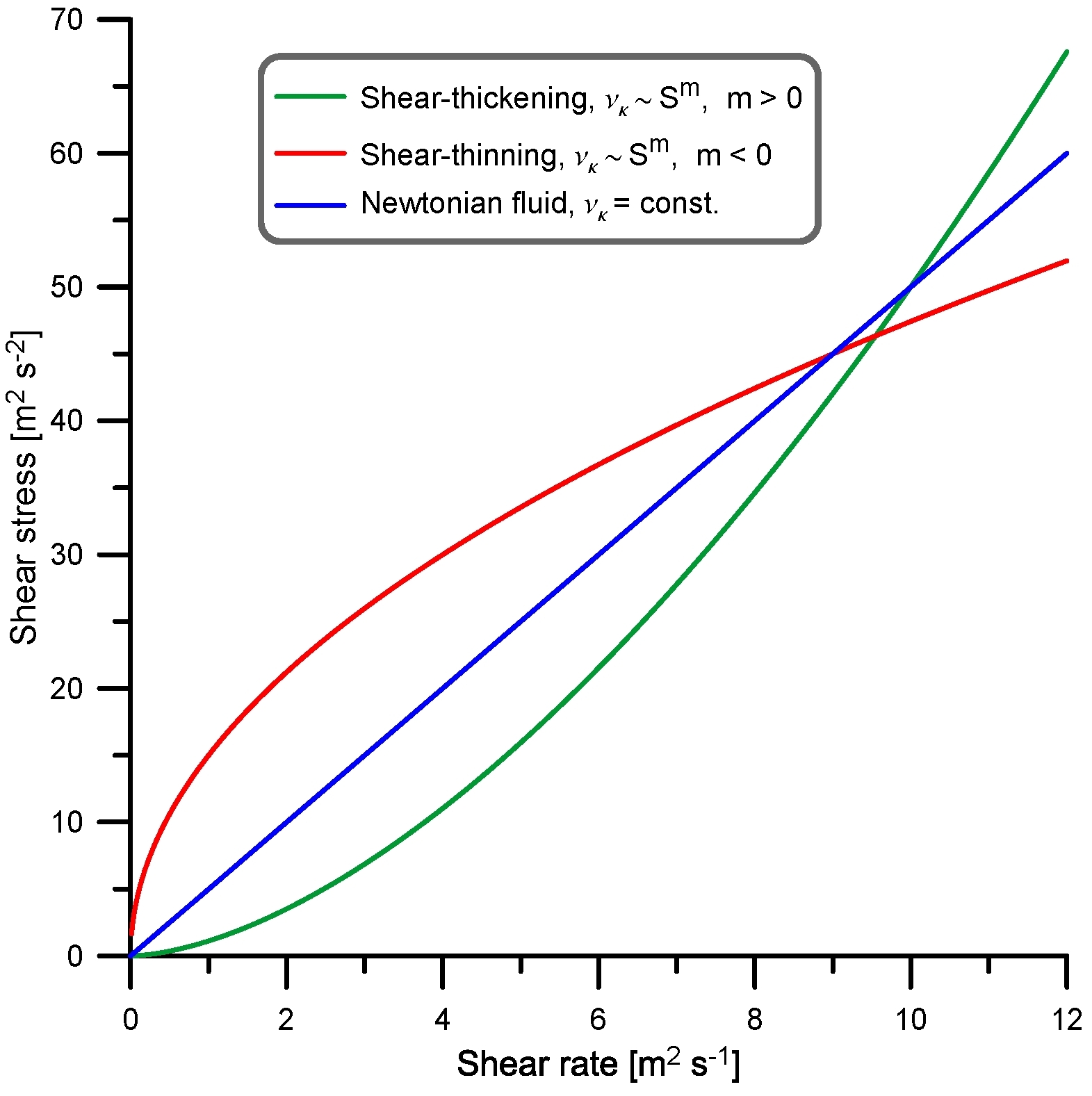}
  \caption{Examples of shear behavior in a shear-thickening (green, m=0.65), a shear-thinning (red) and a Newtonian fluid.}
  \label{kine1}
\end{figure}
The kinematic viscosity of many non-Newtonian fluids may be described 
as a power-law function of local shear rate, $S$, by the following slightly generalized \textit{Ostwald-de Waehle} ansatz, wherein
$\nu_0=\nu_0(T)$ and $\gamma=\gamma(T)$  are (generally temperature-dependent) empirical parameters:
\begin{equation}\label{ostwald}
	\nu = \nu_0 + \gamma\times ({S}/{S_*})^m\,.
\end{equation}
Fig.\ \ref{kinemat} shows three example fluids with (not completely fictive) parameters like 
$\gamma=10^{-6}$ m$^2$\,s$^{-1}$, $S_*=0.19$ s$^{-1}$ (Hz), $\nu_0=0.43$ m$^2$\,s$^{-1}$, and with $m=\{-0.5; +0.65; +1.5\}$. 

In the following we exclusively deal with \textit{shear-thickening fluids} ($m > 0$) and use the notion of turbulence in a narrow sense 
defined later. Nevertheless the theoretical foundations cover a broader range of applications than only the shear-thickening case.

\subsection*{Theoretical foundation\label{found}}

This paper rests essentially on \cite{baumert2013} who 
demonstrated that the  turbulent viscosity, $\nu_t$ [m$^2$\,s$^{-1}$], scaling linearly with turbulent diffusivity, $\mu_t$ [m$^2$\,s$^{-1}$],
may be written as follows,
\begin{equation}\label{eVisc}
	\nu_t= \frac{\cal K}{\pi \,\Omega}
\end{equation}
where $\cal K$ is turbulent kinetic energy in the narrow sense, TKE  [m$^2$\,s$^{-2}$], $\Omega$ is the r.m.s. turbulent vorticity [s$^{-1}$] (a relative of enstrophy) and $\pi$ is
the dimensionless circle number. This formula results from a stochastic-geometric theory of inviscid turbulence wherein turbulent eddies are taken
as singular solutions of the Euler equation and thus as particles (vortex-tubes dipoles) moving in a stochastic fashion -- just like molecules in Einstein's theory of Brownian motion -- 
slightly generalized in form of Fokker-Planck equations wtih generation and dissipation terms. Another result of this theory is the Karman constant as
\begin{equation} \label{karman}
	{\ae}=1{/}\sqrt{2\pi}= 0.40\;.
\end{equation}
\cite[A recent study based on the Princeton Superpipe gave $0.40 \pm 0.02$, see][]{baileyetal2014}.


\section{Inviscid and viscous spectra}\label{spectra}

\subsection*{The major length scales \label{scales}}

Wavenumber spectra of developed viscous turbulence exhibit two distuingished scales: the larger, 
so-called energy-containing scale which depends on the forcing mechanism, $\Lambda$, and the smaller
Kolmogorov scale, $\lambda \ll \Lambda$ , which is related with the kinematic viscosity, $\nu$:
\begin{equation}\label{sKolm}
	\lambda= \left( \frac{ \nu^3 } {\varepsilon}\right)^{1/4 }\;.
\end{equation}
Here $\varepsilon$   [m$^2$\,s$^{-3}$] is the total\footnote{I.e. the integral over the turbulent dissipation spectrum.}  
turbulent dissipation rate of TKE, and $\nu$ [m$^2$\,s$^{-1}$] is the \textit{kinematic} viscosity.
Obviously,  if $\nu\rightarrow 0$, also $\lambda \rightarrow 0$.

The energy-containing turbulent scale, $\Lambda$, has been 
derived on theoretical grounds as follows \cite[]{baumert2013}, 
\begin{equation}\label{Lambda}
	\Lambda= \frac{\sqrt{{\cal K}/\pi}}{\Omega}\,.
\end{equation}
This scale is only indirectly influenced by viscosity.
\begin{figure}[htbp] 
  \includegraphics[width=8.5cm,height=7.cm,keepaspectratio]{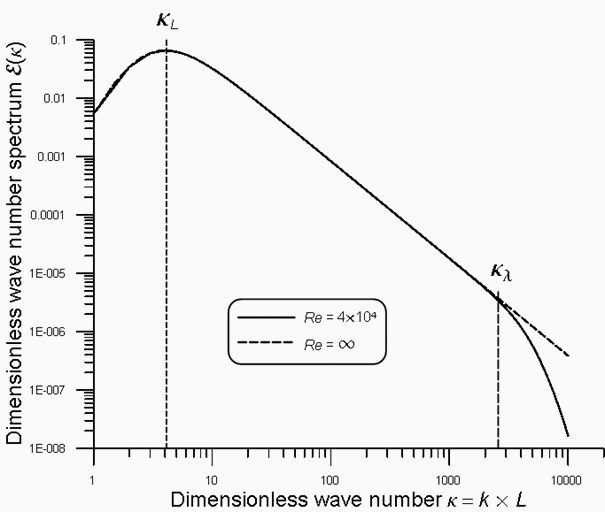}
  \caption{Inviscid (dashed) and viscous (full) spectra of kinetic-energy fluctuations, dimensionless. 
$\kappa_{\Lambda}=2\,\pi$ and $\kappa_{\lambda}=2\,\pi\,{\Lambda}/\lambda$ denote the
dimensionless forms of the energy-containing and the Kolmogorov wavenumber, respectively.}
  \label{spec1}
\end{figure}
\subsection*{Wavenumber spectra\label{spectra}}
For a better understand of our approach we look at Fig. \ref{spec1}. It exhibits a real-world viscous spectrum (full) 
and an idealized spectrum of inviscid turbulence (dashed) discussed in \cite{baumert2013}.
With the dimensionless wave number $\kappa=k\times {\Lambda}$ \cite[eqn. 6.246, p. 232]{pope2000} it reads as follows, 
\begin{equation}\label{essence1}
	{\cal E}(\kappa)= \frac{E(\kappa)}{\varepsilon^{2/3}{\Lambda}^{5/3}} ={C}\times  f_{\Lambda}(\kappa) \times\kappa^{-5/3}\times f_{\lambda}(\kappa, \rho)
\end{equation}
where 
\begin{equation}\label{rho}
	\rho={\Lambda}/\lambda.
\end{equation}
$\rho$ is taken below as a rough proxy of  the Reynols number. 

$C$ in (\ref{essence1}) is a so-called universal Kolmogorov constant for the wavenumber spectrum of turbulence.
It has been derived theoretically as follows \cite[]{baumert2013}:
\begin{equation}\label{KOLMOG1}
	C = \frac{1}{3} (4\pi)^{2/3} = 1.80\,.
\end{equation}
In (\ref{essence1}), $f_{\Lambda}(\kappa)$ describes the low-wavenumber, high-energy interval following Karman's spectral 
model \cite[loc. cit.][p. 747]{pope2000}. $ f_{\lambda}(\kappa, \rho)$ denotes the so-called dissipation region wherein 
the smallest eddy motions are becoming laminar \cite[for details see p. 232, 233 in][]{pope2000}:
\begin{equation}
	f_L(\kappa)=\left(\frac{\kappa}{\sqrt{\kappa^2+c_L}} \right)^{17/3}
\end{equation}
and
\begin{equation}
	f_{\lambda}(\kappa, \rho)= \exp\left[ -\beta \left( \left[ \left(\frac{\kappa}{\rho}\right)^4+c_{\lambda}^4 \right]^{1/4} -c_{\lambda}\right) \right]\, .
\end{equation}
\cite[The exponential ansatz for the dissipation region is in a simpler form due to Kraichnan (1959), loc. cit.][]{pope2000}.
\begin{figure}[tbp] 
  \includegraphics[width=8.5cm,height=5.5cm,keepaspectratio]{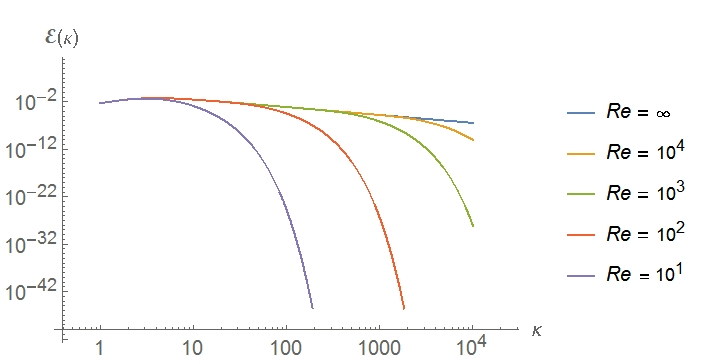}
  \caption{Spectra of kinetic-energy fluctuations for the Reynolds-number proxy $Re=L/\lambda$ decreasing from the upper right curve with $Re=\infty$
down to the lower left with $Re=10$. We see that the area below Kolmogorov's $\kappa^{-5/3}$ inertial subrange  not only 
shrinks quantitatively with decreasing $Re$. It even vanishes completely, indicating the transition from the turbulent to the laminar
state. }
  \label{fig:spec2}
\end{figure}


\section{Turbulence in the narrow sense\\ and other chaotic  motions\label{other}}

In the present paper  we understand the notion \textit{turbulence} in a specific
narrow sense and separate it from other forms of chaotic fluid motions. We  
distinguish three regions of the kinetic-energy wavenumber spectrum. We namely interprete 
Kolmogorov's inertial subrange in the interval [$\kappa_{\Lambda}, \kappa_{\lambda}$]
as \textit{turbulence in the narrow sense} and the rest of the spectrum simply as 
other forms of chaotic fluid motions which mix only weakly.

We differentiate between the following three wavenumber intervals:
\begin{itemize}
  \item[(i)] $\kappa \in [0, \kappa_{\Lambda}]$: weakly mixing Karman range,
  \item[(ii)]  $\kappa \in [\kappa_{\lambda},\kappa_{\Lambda}]$: intensely mixing inertial\\ subrange of Kolmogorov,
  \item[(iii)] $\kappa \in [\kappa_{\lambda},\infty]$: weakly mixing Kraichnan region.
\end{itemize} 
Arguments for (ii) are given in the next section.

Regarding mixing of momentum in the sense of (\ref{eVisc}), only the Kolmogorov range matters which
in the  \textit{inviscid} case ($k_{\lambda}\rightarrow \infty$) extends from the energy-containing scale, $k_{\Lambda}$, until $k_{\lambda}=\infty$ so that 
\begin{equation}\label{tke1a}
	{\cal K}= \int_{k_{\Lambda}}^{\infty }E(k)\,dk=C\times \varepsilon^{2/3} \times \int_{k_{\Lambda}}^{\infty}\frac{dk}{k^{5/3}}\;.
\end{equation}
This last integral gives together with (\ref{Lambda}) a simple analytical result:
\begin{equation}\label{tke1b}
	{\cal K}= C\times \frac{3}{2} \times \left( \frac{\varepsilon\,\Lambda}{2\pi} \right)^{2/3}\;.
\end{equation}
This result has successfully been tested against high-$Re$ turbulence observations
 \cite[][Fig.\ 6]{baumert2013}.

Now we apply this approach to viscous turbulence and write
\begin{equation}\label{tke2a}
	{\cal K}= \int_{k_{\Lambda}}^{k_{\lambda}}E(k)\,dk\;.
\end{equation}
Clearly, (\ref{tke1a}) is a special case of (\ref{tke2a}).

(\ref{tke2a}) is analytically integrable and gives the following:
\begin{equation}\label{tke2b}
	{\cal K}= C \times  \frac{3}{2} \times \left(\frac{\varepsilon}{2\pi}\right)^{2/3} \times \left[ {\Lambda}^{2/3}-{\lambda}^{2/3} \right].
\end{equation}
Due to (\ref{sKolm}), a kinematic viscosity $\nu$ increasing without bounds gives an increasing $\lambda$ until $\lambda = \Lambda$
at which point according to (\ref{tke2b}) TKE vanishes. This means that turbulence in the narrow sense (Kolmogorov's inertial subrange)
disappears and chaotic laminar motions of the now joined Karman-Kraichnan range remain. 
If ${\cal K}=0$, then eddy viscosity $\nu_t$ via (\ref{eVisc}) and consequently also eddy diffusivity $\mu_t$ vanish accordingly.


\section{Mixing within and outside the Kolmogorov subrange}\label{tMix}

Our view of Kolmogorov's inertial subrange has
been filled with life through reports on a series of numerical simulation studies 
\cite[]{herrmann90,herrmannetal90,mannahermann1991} wherein
spectra of space-filling bearing were shown to correspond closely to 
Kolmogorov's 5/3 spectral law.
A space-filling bearing is identical with the densest non-overlapping 
(Apollonian) circle packing in a plane, with side condition that the circles 
are pointwise in contact but able to rotate freely, without friction or slipping
\cite[a 'devil's gear' sensu ][]{poeppe04}. The contact condition for
two different 'wheels' with indices 1 and 2 of the gear reads
\begin{equation}
	\omega_1\times r_1 = \omega_2\times r_2. 
\end{equation}
A \textit{devil's gear} forms spontaneously when 
vortex dipoles collide. Those dipoles move with their inctrinsic velocity governed by
their given radius $r$ and rotation speed $\omega$ \cite[]{baumert2013}. When they
collide they are either scattered or annihilated. The latter case is realized in form of 
a dissipative patch wherein all scales $>0$ are frictionless  except the scale zero
where all the dissipation takes place. Because a dissipative patch in form of 
a \textit{devil's gear} remains at rest, scatter motions scaling with the intrinsic 
parameters $r, \omega$ are the only source of place changes and thus of mixing
\cite[]{baumert2013}. 

Although in these patches scalar mixing takes also place, the bulk of it is due to the place 
changes of dipoles. Hoever, momentum mixing is exclusively due to place changes of 
the dipoles. This is the deeper reason why generally $\mu_t > \nu_t$. I.e. the turbulent 
Prandtl number $\sigma=\nu_t/\mu_t$ is generally less than unity. 


\section{Balance equations for $\cal K$ and $\Omega$\\ in a homgeneously stirred reactor \label{balance}}

\subsection*{General\label{general}}

The dynamic balances of  $\cal K$ and $\Omega$ in a homgeneously stirred reactor are given as follows \cite[]{baumert2013}:
\begin{eqnarray}\label{tke-dyn}
	\frac{d\cal K}{dt} &=& \nu_t \, {S}^2- \varepsilon_h - \varepsilon_b\,,\\
	\frac{d\Omega}{dt} &=& \frac{1}{\pi} \left( \frac{1}{2}{S}^2- \Omega^2\right).\label{om-dyn}
\end{eqnarray}
Here $\nu_t$ is the eddy viscosity given by (\ref{eVisc}), $S=\langle S_e^2\rangle^{1/2}$ 
is the (steady) r.m.s. effective shear rate, $\varepsilon_h={\cal K}\,\Omega/\pi$ \cite[]{baumert05a} is the dissipation 
rate within the bulk volume, and $\varepsilon_b$ is the bulk dissipation within the boundary layers (walls etc.) of the reactor.

In the following we denote steady state values ($d{\cal K}/dt=d\Omega/dt=0$) by an overbar. In particular we have 
\begin{equation}\label{epsb}
	\overline{\varepsilon}_b=\overline{\varepsilon}_h = \frac{\overline{\cal K}\,\overline{\Omega}}{\pi}\,,
\end{equation}
fully indepenent on the forcing ${S}$. Whereas the steady-state case of the dynamic energy balance (\ref{tke-dyn}) 
is useless for the following, the steady-state r.m.s. vorticity balance finds application further below: 
\begin{equation}\label{omss}
	\overline{\Omega}= {S}/\sqrt{2}\,,
\end{equation}

\subsection*{Steady state\label{steady}}

Until now we have no equation yet which gives us the size of Kolmogorov's microscale, $\lambda$. 
It is derived now, based on the above results and thoughts. We begin with (\ref{tke2b}) and replace there
the homogeneous volume dissipation rate via $\overline{\varepsilon} =\overline{\varepsilon}_h= \overline{\cal K}\;\overline{\Omega}/\pi$ 
and find with (\ref{omss}) the following relation for the steady-state turbulent kinetic energy in the reactor:
\begin{eqnarray}\label{kkk}
	\overline{\cal K}&=& \alpha_1 \times \overline{S}^2\times \left[ {\Lambda}^{2/3}-{\overline{\lambda}}^{2/3} \right]^3 \,,\\
	\alpha_1&=& (2\pi)^{-2}\;\;\; =\;\;\; 0.0253.\label{a1}
\end{eqnarray}
This result allows to reformulate the steady-state eddy viscosity (\ref{eVisc}) in the reactor as follows:
\begin{eqnarray}\label{nnn}
	\overline{\nu}_t&=& \frac{\overline{\cal K}}{\pi\, \overline{\Omega}} \;\;=\;\;
		\alpha_2\times  {S} \times \left[ {\Lambda}^{2/3}-\overline{\lambda}^{2/3} \right]^3,\\
	\alpha_2&=& 2^{-3/2}\times\pi^{-3}\;\;\; =\;\;\; 0.0114.\label{a2}
\end{eqnarray}
Analogously we may derive the steady-state TKE dissipation rate in the reactor as 
\begin{eqnarray}\label{eee}
	\overline{\varepsilon}&=& \frac{\overline{\cal K}\,\overline{\Omega}}{\pi} \;\; =\;\;
		\alpha_3 \times  {S}^3 \times \left[ {\Lambda}^{2/3}-\overline{\lambda}^{2/3} \right]^3,\\
	\alpha_3&=& 2^{-5/2}\times \pi^{-3}\;\;\; =\;\;\; 0.00570.\label{a3}
\end{eqnarray}


\section{Microscale bifurcation\label{bif}}

\subsection*{Governing equation\label{gov}}

We now consider according to (\ref{sKolm}) the steady-state microscale, $\overline{\lambda}$ and replace 
therein $\varepsilon$ with the reactor's steady-state value (\ref{eee}):  
\begin{equation}\label{lam1}
	\overline{\lambda }= \left(\frac{ \overline{\nu}^3}{\overline{\varepsilon} }\right)^{1/4}
	=\alpha_4\times \left( \frac{\overline{\nu}/{S} }{\Lambda^{2/3}-\overline{\lambda}^{2/3}} \right)^{3/4},
\end{equation}
\begin{equation}
	\alpha_4 = \alpha_3^{-1/4}=1.1509\,.\label{a4}
\end{equation}
For an easy treatment of (\ref{lam1}) we multiply both sides with $\left[ \Lambda^{2/3}-\overline{\lambda}^{2/3} \right]^{3/4}$ 
and get with some algebra and using the abbreviation $\xi=\overline{\lambda}^2$ the following working equation:
\begin{equation}\label{work}
		(\xi\, \Lambda)^{2/3}  =\xi +\alpha_4^{4/3} \, \overline{\nu} /{S} \,.
\end{equation}

\begin{figure}[h] 
  \includegraphics[width=8.5cm,height=5.5cm,keepaspectratio]{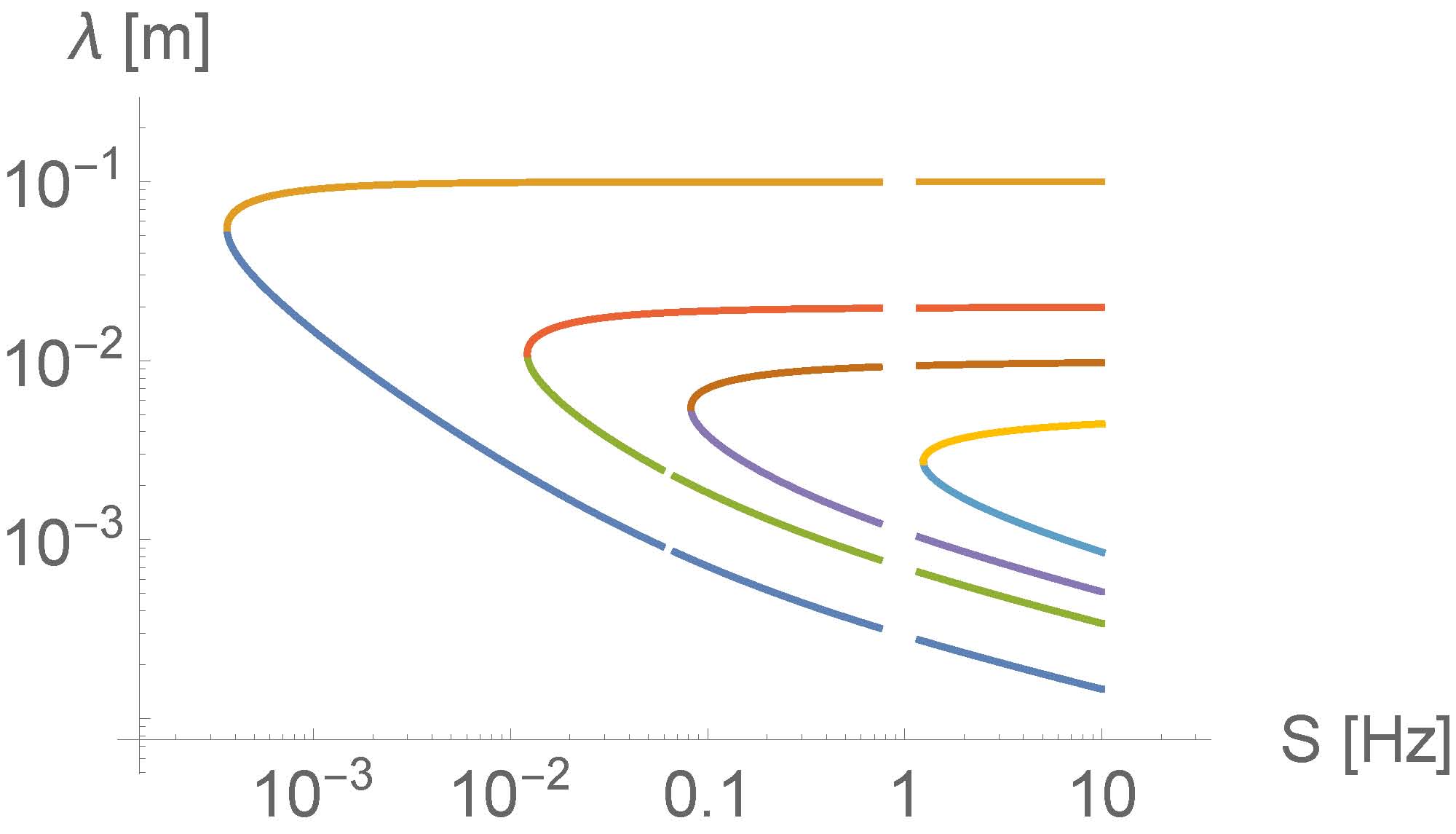}
  \caption{Valid steady-state combinations of forcing parameters {$\Lambda, \lambda$}.
Note that always two states are possible, the upper branch without turbulence in the narrow sense, 
and the lower branch with a developed Kolmogorov spectrum with the property $\lambda \ll \Lambda$.
The forcing length scales are here $\Lambda$ = 10 cm, 2 cm, 1 cm, 5 mm, respectively, from outside to inside.}
  \label{fig:Bild1}
\end{figure}
\begin{figure}[h] 
  \includegraphics[width=8.5cm,height=5.5cm,keepaspectratio]{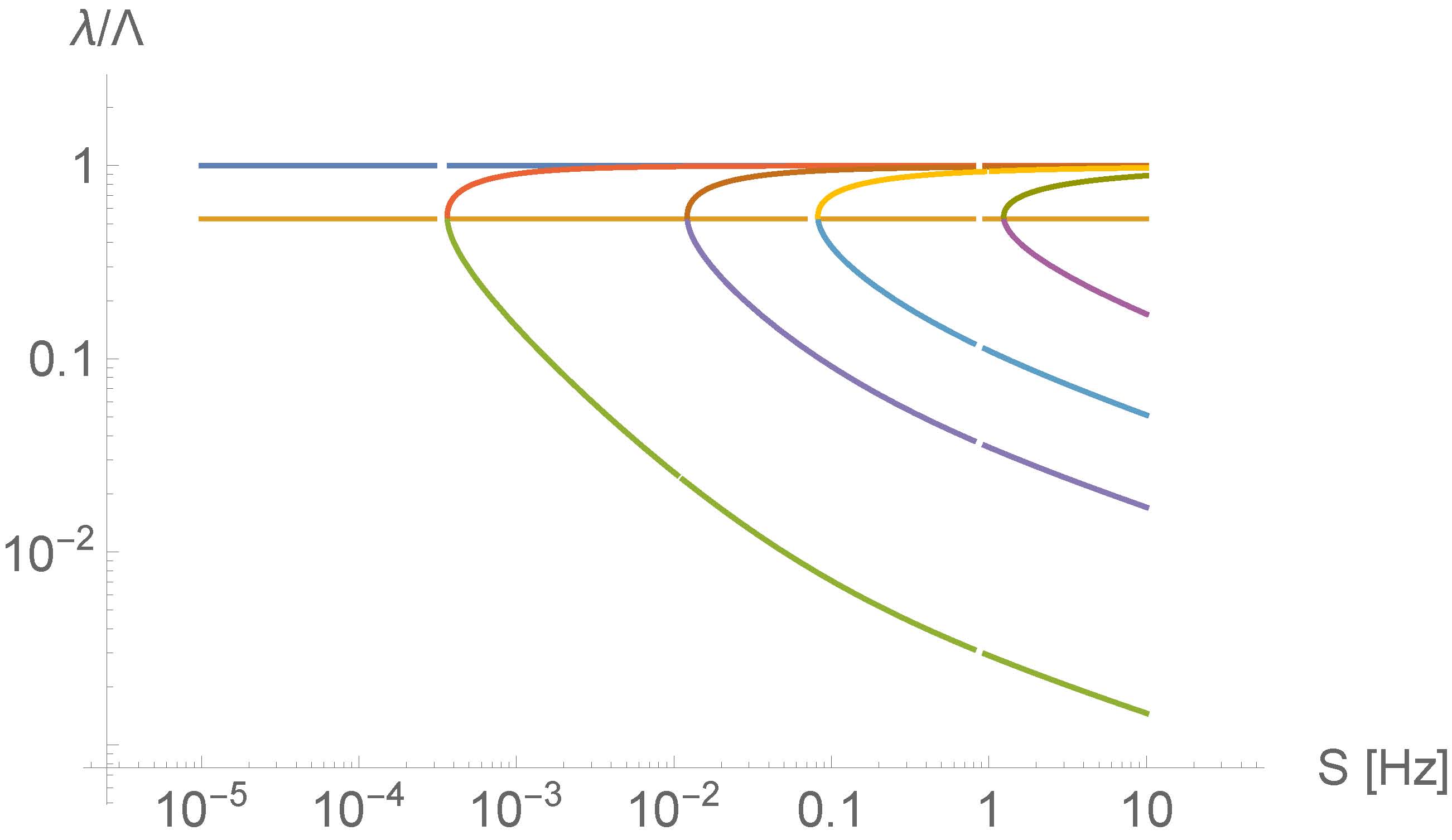}
  \caption{Similar to Fig.\ \ref{fig:Bild1} but instead of $\lambda$ the scale ratio $1/\rho=\lambda/\Lambda$ is shown. 
$\rho$ proper is a proxy for the Reynolds number. $\rho=1$ means that the Kolmogorov scale equals the energy-containing scale so that
the highly mixing inertial subrange of the fluctuation spectrum vanishes and mixing breaks more or less down.
The forcing length scales are here again $\Lambda$ = 10 cm, 2 cm, 1 cm, 5 mm (from outside to inside). Note that the second horizontal
line is situated at $\lambda/\Lambda \approx 0.53$ and labels the border between the upper and lower solution branches of (\ref{work}). }
  \label{fig:Bild2}
\end{figure}


\subsection*{Physical interpretation\label{interpret}}

The meaning of (\ref{lam1}) is most easily understood by highlighting \textit{input} and \textit{output} variables
and assuming mechanical stirring: 
\begin{itemize}
  \item The shear frequency, $\mathbf S$, is controlled by the rotation speed of (e.g.)  the propeller(s) used to stir the reactor. I.e.\ 
		for the solution of (\ref{lam1}),  $\mathbf S$ is thus a { given}, { prescribed \bf input} quantity.
  \item The turbulent length scale $\mathbf \Lambda$ is controlled by geometry (e.g.\ radius, length of stirring propeller etc.) and represents also 
		a {given}, prescribed {\bf input} quantity.
\item The microscale $\overline{\mathbf \lambda}$ however is not given. It is the dynamic relaxation {\bf result} of the forcing and thus an {\bf output} quantity.
\end{itemize}
As we see in Figures \ref{fig:Bild1} and \ref{fig:Bild2}, for a given energy-containing length scale $\Lambda$ (i.e. along a chosen curve), 
each admissible shear value $S$ allows for \textbf{two} values of Kolmogorov's microscale, $\lambda$: 
\begin{itemize}
  \item[a)] the upper value is associated with the strangling of turbulence due to the close proximity of $\lambda$ and $\Lambda$.
This proximity leaves not enough space for a developed turbulence spectrum (in the narrow sense from above). 
\item [b)] The (mostly much) lower value is associated with a good mixing behavior of the stirred reactor. 
Unfortunately nature gives no stability guarantee for this state because there exists the ``official'' alternative of a strangeled 
spectrum -- in agreement with the law. Some aspects are discussed in the next Section \ref{dynprob}. 
\end{itemize}


\section{The full dynamic problem\label{dynprob}}

\subsection*{Smart fluids, smart control?\label{general2}}

We have shown in Section \ref{bif} that a reactor holding  a shear-thickening fluid may attain  
two different steady states with respect to scalar mixing and turbulence. One of them  mixes well, the other one does not. 

Clearly the system cannot stay in both states at the same time. But  non-steady
transitions between them are physically possible and of practical interest. In Figures \ref{fig:Bild1} and \ref{fig:Bild2} these 
transitions occur ouside the computed curves. 

Questions remain. E.g. how does the system choose, for a given set of initial conditions $\{ {\cal K}_0$, $\Omega_0\}$ and forcing parameters $\{ \Lambda, S\}$,
between the two admissible steady states, and under which conditions (including initialization) transitions between them occur.
Possibly smart fluids deserve smart means to keep them under control. 

\subsection*{Relaxation of the ``inner fluid''\label{relax}}

Due to the inner structure of shear-thickening colloidal systems we have to expect that a viscosity according to (\ref{eVisc}) is
never instantaneously realized. Dynamic adjustments to a new micro-mechanical state take certain relaxation steps. 
They are governed by the laws of irreversible thermodynamics \cite[]{wessling1991,wessling1993,wessling1995}. 
Therefore, to get a complete \textit{dynamic} picture of the stirred fluid with our focus on learning \textit{how} one of the two steady states
is selected and \textit{how}  potential transitions between them are controllable, we  actually have at least to augment the equation system (\ref{tke-dyn}, \ref{om-dyn}) 
by an additional equation for the dynamic relaxation of the kinematic viscosity (the ``inner fluid'') against instantateous changes in the 
shear rate, $S=S(t)$. 

These results now quantitatively explain the previously only
 empirically observed non-Newtonian viscosity due to structure build-up
 and -breakdown in colloidal systems under different shear conditions.
They are fully in line with  thermodynamical non-equilibrium
 character of colloidal systems: the result of chaotic dispersion
 processes (which are empirically known not to be easy to control and
 having quite often unexpected results which require intensive studies
 of the process parameters until finally in mass production reproducible
 results can be achieved)

\subsection*{Coda}

At least in the present moment we have no detailed knowledge of  the inner relaxation behavior of our test fluid.
Therefore we have to apply here the most simple ansatz -- linear or ``first order'' relaxation (``nudging''). 
We  augment (\ref{tke-dyn}, \ref{om-dyn}) as follows, 
\begin{eqnarray}
	\label{nu-1}
	\frac{d\nu}{dt}&=& \frac{1}{\tau} \left(\tilde{\nu} - \nu \right).
\end{eqnarray}
This equation describes the inner microscopic-dynamic relaxation of $\nu=\nu(t)$ wherein 
$\tilde\nu$ is the steady-state value of the kinematic viscosity and known from above
as the \textit{Ostwald-de Waehle} relation,
\begin{equation}\label{ostwald2}
	\tilde\nu = \nu_0 + \gamma\times ({S}/{S_*})^m\,.
\end{equation}
Thus finally we have
\begin{eqnarray}
	\label{nu-2}
	\frac{d\nu}{dt}&=& \frac{1}{\tau} \left[ \nu_0 + \gamma\times \left(\frac{S}{S_*}\right)^m - \nu \right].
\end{eqnarray}
This is our first attempt to describe the effects of changes in nanoscopic structures of colloidal systems with respect to viscosity.

The value of $\tau$ can  be measured in principle but is not yet available. 
Its value is important because its relation to intrinsic turbulent time scales matter. 
It is also not clear whether the linear ansatz is the correct one among the many other relaxation models.
Further, although  (\ref{tke-dyn}, \ref{om-dyn}, \ref{nu-2}) form a closed dynamical problem, 
(\ref{tke-dyn},\ref{om-dyn}) are possibly not sufficiently representative for the state of strangled 
eddy viscosity \cite[they have been derived for \textit{inviscid} flows,][]{baumert2013}.
These questions deserve specific experimental studies and further theoretical efforts. 

{\bf Acknowledgements.} HZB thanks \textit{Alexander J. Babchin} in Tel Aviv for dropping hints on 
early fruitful efforts of \textit{Yakov Frenkel} (1894 -- 1952) towards a kinetic theory of liquids and 
certain generalized phase transitions. 


\nocite{frenkel1946}
\bibliography{nonNewtonian,bib1,bib2,bib3,bib4}

\begin{thebibliography}{12}
\expandafter\ifx\csname natexlab\endcsname\relax\def\natexlab#1{#1}\fi
\expandafter\ifx\csname url\endcsname\relax
  \def\url#1{{\tt #1}}\fi
\expandafter\ifx\csname urlprefix\endcsname\relax\def\urlprefix{URL }\fi
\expandafter\ifx\csname doiprefix\endcsname\relax\def\doiprefix{doi:}\fi

\bibitem[{Bailey et~al.(2014)Bailey, Vallikivi, Hultmark, and
  Smits}]{baileyetal2014}
Bailey, S. C.~C., M.~Vallikivi, M.~Hultmark, and A.~J. Smits, 2014: Estimating
  the value of von {K}arman's constant in turbulent pipe flow. {\it J. Fluid
  Mech.\/}, {\bf 749}, 79 -- 98, doi:dx.doi.org/10.1017/jfm.2014.208.

\bibitem[{Baumert(2005)}]{baumert05a}
Baumert, H.~Z.: 2005, A novel two-equation turbulence closure for high
  {R}eynolds numbers. {Part B: S}patially non-uniform conditions. {\it Marine
  Turbulence: Theories, Observations and Models\/}, H.~Z. Baumert, J.~H.
  Simpson, and J.~S{\"u}ndermann, eds., Cambridge University Press, Chapter 4,
  31 -- 43.

\bibitem[{Baumert(2013)}]{baumert2013}
Baumert, H.~Z., 2013: Universal equations and constants of turbulent motion.
  {\it Physica Scripta\/}, {\bf T155}, 014001 (12pp),
  doi:10.1088/0031-8949/2013/T155/014001.

\bibitem[{Frenkel(1946)}]{frenkel1946}
Frenkel, Y., 1946: {\it Kinetic Theory of Liquids\/}. Clarendon Press, 1st
  engl. ed., 488 pp., russian ed. 1943, Kazan.

\bibitem[{Herrmann(1990)}]{herrmann90}
Herrmann, H.~J.: 1990, Space-filling bearings. {\it Correlation and
  Connectivity\/}, H.~E. Stanley and N.~Ostrowsky, eds., Kluwer Academic Publ.,
  108 -- 120.

\bibitem[{Herrmann et~al.(1990)Herrmann, Mantica, and Bessis}]{herrmannetal90}
Herrmann, H.~J., G.~Mantica, and D.~Bessis, 1990: Space-filling bearings. {\it
  Phys. Rev. Letters\/}, {\bf 65}, 24 December, 3,223 -- 3,226.

\bibitem[{Manna and Herrmann(1991)}]{mannahermann1991}
Manna, S.~S. and H.~J. Herrmann, 1991: Precise determination of the fractal
  dimensions of apollonian packing and space-filling bearings. {\it Journal of
  Physics A\/}, {\bf 24}.

\bibitem[{Pope(2000)}]{pope2000}
Pope, S.~B., 2000: {\it Turbulent flows\/}. Cambridge U. Press, 771 pp.

\bibitem[{P\"oppe(2004)}]{poeppe04}
P\"oppe, C., 2004: {Das Getriebe des Teufels}. {\it Spektrum d. Wiss.\/},
  September, 104 -- 109.

\bibitem[{Wessling(1991)}]{wessling1991}
Wessling, B., 1991: Dispersion hypothesis and non-equilibrium thermodynamics:
  key elements for a materials science of conductive polymers. {A key to
  understanding polymer blends or other multiphase polymer systems}. {\it
  Synthetic metals\/}, {\bf 45}, 119 -- 149.

\bibitem[{Wessling(1993)}]{wessling1993}
--- 1993: Dissipative structure formation in collodial systems. {\it Advanced
  Materials\/}, {\bf 5}, 300 -- 305.

\bibitem[{Wessling(1995)}]{wessling1995}
--- 1995: Critical shear rate -- the instability reason for the creation of
  dissipative structures in polymers. {\it Z. Phys. Chem.\/}, {\bf 191}, 119 --
  135.

\end{thebibliography}
\end{document}